\documentclass{iopart}
\usepackage{epsf,amssymb,psfrag,iopams,setstack}%,amsmath}
\usepackage{graphicx}
\catcode`\@=11
\def \theequation {\thesection.\arabic{equation}}

\def \be  {\begin{equation}}
\def \ee  {\end{equation}}
\def \ba  {\begin{eqnarray}}
\def \ea  {\end{eqnarray}}
\def \dd  {\partial}

\def \matrix #1 {\left(\begin{array}{cc} #1 \end{array}\right)}

\def \tr {\mathop{\rm tr}\nolimits}

\def \Re {\mathop{\rm Re}\nolimits}

\newcommand \widebar [1] {\overline{#1}}

\newcommand{\as}{\ifmmode\alpha_{\rm s}\else{$\alpha_{\rm s}$}\fi}
\newcommand{\asbar}{\ifmmode\bar{\alpha}_{\rm s}\else{$\bar{\alpha}_{\rm s}$}\fi}

\def \CR {{\mathcal R}}

\def \VV {\mathbf V}

\def\inbar{\,\vrule height1.5ex width.4pt depth0pt}
\def\IC{\relax\hbox{$\inbar\kern-.3em{\rm C}$}}
\def\IZ{\relax{\hbox{\cmss Z\kern-.4em Z}}}
\def\IR{{\hbox{{\rm I}\kern-.2em\hbox{\rm R}}}}

\def\IP{{\hbox{{\rm I}\kern-.2em\hbox{\rm P}}}}
\def\II{\hbox{{1}\kern-.25em\hbox{l}}}

\def\numberbysection{\@addtoreset{equation}{section}
                     \def\theequation{\thesection\arabic{equation}}}
\newcommand \Mybf[1] {\mbox{\boldmath$ {#1} $}}
\newbox\lett\newdimen\lheight\newdimen\lwidth
\def\ontop#1#2{\setbox\lett=\hbox{#2}\lheight\ht\lett
\multiply\lheight by 12 \divide\lheight by 10\relax%
\lwidth\wd\lett \multiply\lwidth by 8 \divide\lwidth by 10\relax #2\kern-\lwidth%
\raise\lheight\hbox{{$\scriptstyle #1$}}\kern.1ex}

\begin{document}
\title[Factorization of the transfer matrices]{Factorization
of the transfer matrices for the quantum $s\ell(2)$ spin
chains and Baxter equation}
\author{S {\'E} Derkachov~$^1$,~~ A N Manashov~$^2$\footnote{ Permanent
address:\ Department of Theoretical Physics,  Saint-Petersburg State University,
\mbox{St.-Petersburg,} Russia}}

\address{$^1$ St.Petersburg Department of Steklov
Mathematical Institute of Russian Academy of Sciences,
Fontanka 27, 191023 St.Petersburg, Russia.
 }

\address{$^2$ Institute for Theoretical Physics, University of  Regensburg,\\
D-93040 Regensburg, Germany}

\begin{abstract}
It is shown that the transfer matrices of homogeneous $s\ell(2)$ invariant spin chains with
generic spin, both closed and open, are factorized
into the product of two operators. The latter satisfy the Baxter equation that follows
from the structure of the reducible representations of the $s\ell(2)$ algebra.
\end{abstract}
\maketitle

\setcounter{footnote}{0}
\section{Introduction}
The recent interest to the analysis of  noncompact spin magnets (spin chains with
the infinite dimensional Hilbert space at each site) is motivated by the advances in gauge
field theories (see for a review~\cite{BBGK,AT}). These models (spin magnets) can be solved
with the help of the
Algebraic Bethe Ansatz (ABA) method~\cite{FST,LF}.
Alternatively, the solution
is provided by the method of the Baxter $Q-$operators~\cite{Baxter}.

The Baxter $Q-$operator is known for a large number of 
integrable
models~\cite{PG,V95,BLZ,SD,Pronko,KSS,RW,open,KM,Korff1}.
Nevertheless, a universal method for obtaining of the Baxter
operator is absent so far and each model (or class of
models) needs a special consideration. The derivation of
the Baxter $Q-$operator for the $s\ell(2)$ spin chain
models is based on the Pasquier-Gaudin trick, see
Refs.~\cite{PG,SD,DKM01}. The generalization of the latter
to the spin chains with the higher rank symmetry groups is
not quite obvious.

In the present paper we give the alternative derivation of
the Baxter equation for the noncompact $XXX$ spin chain
models. We shall show that the transfer matrices for the
homogeneous spin chain models  factorize into a product of
two operators. The factorization holds for all closed
$s\ell(2)$ spin chains studied so far~\cite{DKM01,KM,SD2}
and can be traced to the factorization of the $\mathcal
R-$operator obtained in Ref.~\cite{SD1}. We prove that this
property is true for the open spin chain models as well. As
the factorization property is established, the Baxter
equation for these operators can be deduced in a simple way
from the structure of the reducible representations of
the $s\ell(2)$ algebra. (See  Ref.~\cite{AF} where similar
arguments were applied to the analysis of $q-$deformed spin
chain models.) We shall consider the spin chains with the
quantum space being the generic lowest weight
representation of the $s\ell(2)$ algebra, but the method
works  for the principal series representations of the
$SL(2,{\mathbb R})$  (\,$SL(2,{\mathbb C})$\,) group as well.
Taking into account that, as it was shown in Ref.~\cite{SD1},
factorization holds for  the $s\ell(3)$ and
$s\ell(2|1)$ invariant $\mathcal R-$operators one can hope
that the  approach presented here admits a generalization
for the spin chains with the symmetry group of  higher
rank.

The paper is organized as follows: In 
Section~\ref{prelim} we introduce notations and describe
the model. In  Section~\ref{fact} we prove the
factorization property for the transfer matrices for both
the  closed and open $s\ell(2)$ noncompact spin chain
models. In  Section~\ref{BAX} the derivation of the
Baxter equation based on the structure of the reducible
$s\ell(2)$ representations is given. 
Section~\ref{Concl} contains concluding remarks.

%%%%%%%%%%%%%%%%%%%%%%%%%%%%%%%%%%%%%%%%%%%%%%%%%%%%%%%%%%%%%%%%%%%%%%%%%%%%%%%%%%
%%%%%%%%%%%%%%%%%%%%%%%%%%%%%%%%%%%%%%%%%%%%%%%%%%%%%%%%%%%%%%%%%%%%%%%%%%%%%%%%%%
\section{Preliminaries}\label{prelim}
%%%%%%%%%%%%%%%%%%%%%%%%%%%%%%%%%%%%%%%%%%%%%%%%%%%%%%%%%%%%%%%%%%%%%%%%%%%%%%%%%%%%%%
%%%%%%%%%%%%%%%%%%%%%%%%%%%%%%%%%%%%%%%%%%%%%%%%%%%%%%%%%%%%%%%%%%%%%%%%%%%%%%%%%%%%%%
The basic object in the theory of the lattice integrable
systems is a $\mathcal R-$operator. The $\mathcal R-$ operator is a linear operator which
depends on a spectral parameter $u$ and  acts on the tensor product of
two $s\ell(2)$ modules (representations of the $s\ell(2)$
algebra). It satisfies the Yang-Baxter relation (YBR)
\begin{equation}\label{YB}
\CR_{12}(u)\CR_{13}(u+v)\CR_{23}(v)=\CR_{23}(v)\CR_{13}(u+v)\CR_{12}(u)\,.
\end{equation}
The operators acts on the tensor product
$\VV_1\otimes\VV_2\otimes\VV_3$, and, as usual, indices $ik$
indicate that the operator $\CR_{ik}$ acts nontrivially on
the tensor product $\VV_i\otimes\VV_k$. We shall consider
the $s\ell(2)$ invariant solutions of the YBR.

The $s\ell(2)$ algebra has three generators $S_+,\,S_-$ and
$S_0$ which satisfy the well known commutation relations
\begin{equation}\label{com}
[S_0,S_\pm]=\pm S_\pm,\ \ \ \ \ [S_+,S_-]=2S_0\,.
\end{equation}
The lowest weight representation of $s\ell(2)$ algebra,
$D_s$, is uniquely determined by the complex number ( spin
) $s$. The generators can be realized as the differential
operators
\begin{equation}\label{gen}
S_-=-\partial_z,\ \ \ S_+=z^2\partial_z+2sz,\ \ \ S_0=s\partial_z+s
\end{equation}
acting on the linear space ${\mathbf V}_s={\mathbb C}[z]$
(the space of polynomials of arbitrary degree of a
complex variable $z$). For a given $s$ the
representation~\eref{gen} is irreducible unless $s$ is a
negative (half)integer. If $s=-n$, $n=0,1/2,1,...$ the
space ${\mathbf V}_s$ contains a finite dimensional
invariant subspace, ${V}_{n}$, the space of polynomials of
degree less or equal to $2n$, ($\dim V_{n}=2n+1$). The
representation induced on the factor space $\VV_{-n}/V_{n}$
is equivalent to the representation $D_{s'}$ with spin
$s'=1+n$.  The operator $\mathrm{A}$ which intertwines the
representations $D_{-n}$ and $D_{n+1}$,
($\mathrm{A}\,D_{-n}=D_{1+n} \, \mathrm{A}$) ,  is defined
by the commutation relations
$$
\mathrm{A}\,S_\alpha^{(s=-n)} = S_\alpha^{(s=n+1)} \, \mathrm{A}\
$$
and has the form $\mathrm{A} =\partial_z^{2n+1}$.

For the real $s>1/2$ there exists the invariant scalar product
$(\cdot,\cdot)_s$ on the space $V_s$,
\begin{equation}\label{sc}
(\psi_1,\psi_2)_s=\int {\mathcal D}_sz\,\overline{\psi_1(z)}\,{\psi_2}(z)\,,
\end{equation}
where
\begin{equation}\label{DS}
\int{\mathcal D}_s z \,\varphi(z,\bar z)\equiv \frac{2s-1}{\pi}\,\ \int
_{|z|<1}\,d^2z \,(1-|z|^2)^{2s-2}\, \varphi(z,\bar z) \,.
\end{equation}

The operator $S_0$ is  hermitian with respect to the
scalar product~(\ref{sc}), while $S_-^\dagger=-S_+$. For
complex $s$ the integral~\eref{sc} defines the invariant
bilinear form on the  tensor product $\VV_{s^*}\otimes
\VV_s$. The unit operator (reproducing kernel) has the form
\begin{equation}
{\mathbb K}_s(z,w)=(1-z\bar w)^{-2s}\,.
\end{equation}
The identity
\begin{equation}\label{KK}
\psi(z)=\int {\mathcal D}_s w\, {\mathbb K}_s(z,w)\,\psi(w)\,,
\end{equation}
where $\psi(w)$ is the function analytic in the unit circle
holds for  complex $s$ such that $\Re s>1/2$; for all
other spins it should be understood as an analytic continuation
in $s$. \vskip 0.5cm

The $s\ell(2)$ invariant $\mathcal R-$operator acting on the tensor product of two spaces
$\VV_{s_1}\otimes \VV_{s_2}$ has the form~\cite{PKS,LF,ES91}
\begin{equation}\label{Rop}
{\mathcal R}_{12}(u) =(-1)^{\mathbb
J-s_1-s_2}\frac{\Gamma(s_1+s_2+iu)}{\Gamma(s_1+s_2-iu)}
\frac{\Gamma({\mathbb J}-iu)}{\Gamma({\mathbb J}+iu)}\,,
\end{equation}
where the operator of the conformal spin ${\mathbb J}$ is related to the two-particle Casimir
operator in the standard manner
\begin{equation}
{\mathbb J}({\mathbb J}-1)=(\vec{S}_1+\vec{S}_2)^2\,.
\end{equation}
It was shown in the Ref.~\cite{SD1}  that the ${\mathcal R}-$operator~\eref{Rop} can be
represented in the factorized form
\begin{equation}\label{R}
{\mathcal R}_{12}(u)=P_{12} \,{\mathcal
R}_{12}^+(\alpha)\,{\mathcal R}_{12}^-(\beta)=
P_{12}\,{\mathcal R}_{12}^-(\beta)\,{\mathcal
R}_{12}^+(\alpha)\,.
\end{equation}
Here $P_{12}$ is the permutation operator $P_{12}\psi(z_1,z_2)=\psi(z_2,z_1)$,
and
\begin{equation}\label{ab}
\alpha=\frac{s_2-s_1+iu}{2}\,,\   \  \       \ \ \ \ \beta=\frac{s_1-s_2+iu}{2}\,.
\end{equation}
The operator ${\mathcal R}_{12}^-(\alpha)$ is a
$s\ell(2)$ covariant operator, i.e. it maps
$$
\VV_{s_1}\otimes \VV_{s_2}\to \VV_{s_1-\alpha}\otimes
\VV_{s_2+\alpha}
$$
and has the following form
\begin{equation}
{\mathcal
R}_{12}^-(\alpha)=\frac{\Gamma(2s_1)}{\Gamma(2s_1-2\alpha)}
\frac{\Gamma(z_{12}\partial_1+2s_1-2\alpha)}
{\Gamma(z_{12}\partial_1+2s_1)}\,,
\end{equation}
where $z_{12}=z_1-z_2$. Such normalization implies that
$R_{12}^-(0)={\mathbb I}$ and $R_{12}^-(\alpha)\cdot 1=1$.
The second operator, ${\mathcal R}_{12}^+(\alpha)$,\ 
(${\mathcal R}_{12}^+(\alpha):\,\VV_{s_1}\otimes \VV_{s_2}\to \VV_{s_1+\alpha}\otimes
\VV_{s_2-\alpha} $) is
\begin{equation}
{\mathcal R}_{12}^+(\alpha)={\mathcal
R}_{21}^-(\alpha)=\frac{\Gamma(2s_2)}{\Gamma(2s_2-2\alpha)}
\frac{\Gamma(z_{21}\partial_2+2s_2-2\alpha)}
{\Gamma(z_{21}\partial_2+2s_2)}\,.
\end{equation}
The operators ${\mathcal R}^\pm_{12}(\alpha)$ depend on
three parameters -- the spins $s_1,s_2$ and the
spectral parameter $\alpha$. The spins are always fixed by
the tensor properties of the space $\VV_{s_1}\otimes
\VV_{s_2}$ the operators act on, therefore we shall
display the dependence of the operators on the spectral
parameter only. The action of the $\mathcal
R-$operator~\eref{R} on the space $\VV_{s_1}\otimes
\VV_{s_2}$ results in the following chain of 
transformations
$$
\VV_{s_1}\otimes \VV_{s_2}{\overset{{\mathcal
R_{12}^-(\beta)}}{\longrightarrow}}
\VV_{(s_1+s_2-iu)/2}\otimes \VV_{(s_1+s_2+iu)/2}
{\overset{{\mathcal
R_{12}^+(\alpha)}}{\longrightarrow}}\VV_{s_2}\otimes
\VV_{s_1}
{\overset{P_{12}}{\longrightarrow}}\VV_{s_1}\otimes
\VV_{s_2}\,.
$$
 In the next section we shall represent the operators
${\mathcal R}_{12}^\pm(\alpha)$ as  integral operators
and prove the factorization of the transfer
matrices~\cite{FST,ES88}
\begin{eqnarray}
\label{Tc}
{\mathbf T}^{\mathrm{ cl}}_{s_0}&=&\tr_{s_0}{\mathcal R}_{10}(u)
\ldots {\mathcal R}_{N0}(u)\,,\\
{\mathbf T}^{\mathrm{ op}}_{s_0}&=&\tr_{s_0}{\mathcal
R}_{10}(u)\ldots {\mathcal R}_{N0}(u) {\mathcal
R}_{N0}^{-1}(-u) \ldots{\mathcal R}_{10}^{-1}(-u)
\end{eqnarray}
for the homogeneous closed and open $s\ell(2)$ invariant
spin chains. The $\mathcal R-$operator obeys the relation
${\mathcal R}_{12}^{-1}(u)={\mathcal R}_{12}(-u)$ so that
we shall use the following expression for the ${\mathbf
T}^{\mathrm{ op}}_{s_0}$
\begin{eqnarray}
\label{To}{\mathbf T}^{\mathrm{
op}}_{s_0}&=&\tr_{s_0}{\mathcal R}_{10}(u)\ldots {\mathcal
R}_{N0}(u) {\mathcal R}_{N0}(u) \ldots{\mathcal R}_{10}(u).
\end{eqnarray}

\setcounter{equation}{0}
\section{Factorization}\label{fact}
%%%%%%%%%%%%%%%%%%%%%%%%%%%%%%%%%%%%%%%%%%%%%%%%%%%%%%%%%%%%%%%%%%%%
%%%%%%%%%%%%%%%%%%%%%%%%%%%%%%%%%%%%%%%%%%%%%%%%%%%%%%%%%%%%%%%%%%%%
\begin{figure}[t]
\psfrag{2j1}[cc][cc]{${2s_1-2\alpha}$}
\psfrag{2j2}[cc][cc]{${2s_2}$}
\psfrag{2a}[lc][lc]{${2\balpha}$}
\psfrag{z1}[cc][cc][1.2]{$\bar w_1$}
\psfrag{z2}[cc][cc][1.2]{$\bar w_2$}
\psfrag{w1}[cc][cc][1.2]{$ z_1$} \psfrag{w2}[cc][cc][1.2]{$
z_2$}
%\psfrag{s1}[cc][cc]{$2s_1$}
%\psfrag{s2}[cc][cc]{$2s_2$}
%\centerline{\epsfxsize5.0cm\epsfbox{U.eps}}
\centerline{\includegraphics[scale=0.7]{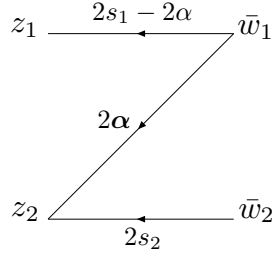}}
%\vspace*{0.5cm}
\caption[]{Graphical representation of the ${\mathcal
R}_{12}^-(\alpha)-$operator. The arrow with the index
$\alpha$ directed from $\bar w$ to $z$ denotes the factor
$(1-z\bar w)^{-\alpha}$.} \label{U}
\end{figure}
%%%%%%%%%%%%%%%%%%%%%%%%%%%%%%%%%%%%%%%%%%%%%%%%%%%%%%%%%%%%%%%%%%%%
%%%%%%%%%%%%%%%%%%%%%%%%%%%%%%%%%%%%%%%%%%%%%%%%%%%%%%%%%%%%%%%%%%%%
We find convenient to represent all operators in question
as integral operators. Let us write the action of the
operator ${\mathcal A}$ on the function $\psi\in
\prod_{k=1}^N\otimes\VV_{s_k}$ in the following form
\begin{equation}\label{A}
[{\mathcal A}\psi](\Mybf{z})=\int \prod_{k=1}^N{\mathcal D}_{s_k} w_k\,
A(\Mybf{z}|\overline{\Mybf{w}})\,
\psi(\Mybf{w})\,,
\end{equation}
where $\Mybf{z}=(z_1,\ldots,z_N)$. It follows from the
definition~\eref{A} and  Eq.~\eref{KK} that the kernel of
the operator $\mathcal A$ can be obtained as follows
\begin{equation}\label{R-K}
A(\Mybf{z}|\overline{\Mybf{w}})={\mathcal
A}\cdot\prod_{k=1}^N (1-z_k\bar w_k)^{-2s_k}\,.
\end{equation}
Here  the operator ${\mathcal A}$ on the r.h.s of
Eq.~\eref{R-K} acts on  $\Mybf{z}$-variables.
% and
%$\overline{\Mybf{w}}$-variables are exterior parameters.}

It is easy to show that the kernel of the operator ${\mathcal
R}_{12}^-(\alpha)$ takes the following form
\begin{equation}\label{Rint}
\fl {R}_{12}^-(\alpha)(z_1,z_2|\bar w_1, \bar
w_2)=(1-z_1\bar w_1)^{-2s_1+2\alpha} (1-z_2\bar
w_1)^{-2\alpha} (1-z_2\bar w_2)^{-2s_2}\,.
\end{equation}
It is convenient to represent the kernel
$R_{12}^-(\alpha)(\Mybf{z}|\overline{\Mybf{w}})$ in the
graphical form. Namely, let us denote the reproducing
kernel ${\mathbb K}_{\alpha}(z,w)= (1-z\bar w)^{-2\alpha}$
by the arrow with the index $2\alpha$ directed from $w$ to
$z$. Then the kernel ${
R}_{12}^-(\alpha)(\Mybf{z}|\overline{\Mybf{w}})$
 is given by the diagram shown in the Figure~\ref{U}.
Similarly,  as follows from Eq.~\eref{R}, the kernel of the
${\mathcal R}_{12}-$operator has the form
\begin{eqnarray}\fl
R_u(z_1,z_2|\bar w_1,\bar w_2)&=&(1-z_2\bar w_1)^{-2\gamma}
\int D_{(s_1+s_2+iu)/2}\zeta \nonumber\\
&&(1-z_1\bar\zeta)^{-2s_1}
(1-z_2\bar \zeta)^{-2\alpha}
(1-\zeta \bar w_1)^{-2\beta}
(1-\zeta \bar w_2)^{-2s_2}
\end{eqnarray}
where $\alpha$ and $\beta$ are defined in  Eq.~\eref{ab} and
$
\gamma=(s_1+s_2-iu)/2.
$

\begin{figure}[t]
\psfrag{A1}[cc][bc]{$c$}%s_1+s_2-iu$}
\psfrag{A2}[rc][rc]{$a$}%s_2-s_1+iu$}
\psfrag{A3}[lc][lc]{$b$}%s_1-s_2+iu$}
\psfrag{z1}[cc][cc]{$z_1$}
\psfrag{z2}[cc][cc]{$z_2$}
\psfrag{w1}[cc][cc]{$\bar w_1$}
\psfrag{w2}[cc][cc]{$\bar w_2$}
\psfrag{s1}[cc][cc]{$2s_1$}
\psfrag{s2}[cc][cc]{$2s_2$}
\psfrag{N}[cc][cc]{$=$}
\centerline{\includegraphics[scale=0.66]{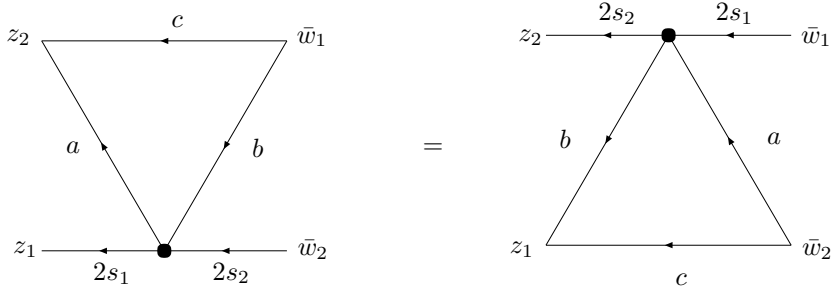}}
\caption[]{The two equivalent graphical representation of the kernel of the $\mathcal R-$
operator. The black dot denotes the integration vertex with the measure
corresponding to the spin $(s_1+s_2+iu)/2$ and the indicies
$a=2\alpha=s_2-s_1+iu$, $b=2\beta=s_1-s_2+iu$, $c=2\gamma=s_1+s_2-iu$.}
\label{RR}
\end{figure}

There exists another equivalent representation for the $\mathcal
R-$operator which follows from the second equality in
Eq.~\eref{R}. Again, it is useful to represent both of them
in the graphical form, see Figure~\ref{RR}.
%%%%%%%%%%%%%%%%%%%%%%%%%%%%%%%%%%%%%%%%%%%%%%%%%%%%%%%%%%%%%%%%%%%%
%%%%%%%%%%%%%%%%%%%%%%%%%%%%%%%%%%%%%%%%%%%%%%%%%%%%%%%%%%%%%%%%%%%%
%%%%%%%%%%%%%%%%%%%%%%%%%%%%%%%%%%%%%%%%%%%%%%%%%%%%%%%%%%%%%%%%%%%%
%%%%%%%%%%%%%%%%%%%%%%%%%%%%%%%%%%%%%%%%%%%%%%%%%%%%%%%%%%%%%%%%%%%%
The identity depicted in  Figure~\ref{RR} (permutation relation) can be considered as
an integral identity between the reproducing kernels. It will be quite useful in the
subsequent analysis.

Let us summarize the properties of the ${\mathcal R}^--$operators.
One easily checks that
\begin{eqnarray}\label{comm}
{\mathcal R}^\pm_{12}(\alpha) \,{\mathcal R}^\pm_{12}(\beta)={\mathcal
R}^\pm_{12}(\alpha+\beta)\,,
\\[2mm]
\label{perm}
{\mathcal R}^+_{12}(s_2-s_1+\alpha)\,{\mathcal R}^-_{12}(\alpha)=
{\mathcal R}^-_{12}(\alpha)\,{\mathcal R}^+_{12}(s_2-s_1+\alpha)\,,
\\[2mm]
\label{F3}
{\mathcal R}^\pm_{12}(\alpha)\,{\mathcal R}^\pm_{23}(\alpha+\beta)\,{\mathcal R}^\pm_{12}(\beta)=
{\mathcal R}^\pm_{23}(\beta)\,{\mathcal R}^\pm_{12}(\alpha+\beta)\,
{\mathcal R}^\pm_{23}(\alpha)\,.
\end{eqnarray}
The first equality follows from  Eq.~\eref{Rint} and from the
property of the reproducing kernel. The second one is the
consequence of  Eq.~\eref{R}. The last one arises as the
selfconsistency relation of the defining equations for the
${\mathcal R}^\pm$ operators~\cite{SD2} and can be 
checked directly by making use of the permutation relation.

Let us introduce operators ${\mathcal
L}^{\pm}_{12}(\alpha)= P_{12}{\mathcal
R}_{12}^{\pm}(\alpha)$. It is straightforward to check that
these operators satisfy the relation
\begin{equation}\label{LB}
{\mathcal L}_{12}^\pm(\alpha){\mathcal L}_{13}^\pm(\alpha+\beta)
{\mathcal L}_{23}^\pm(\beta)={\mathcal L}_{23}^\pm(\beta){\mathcal L}_{13}^\pm(\alpha+\beta)
{\mathcal L}_{12}^\pm(\alpha)\,.
\end{equation}
Equation~\eref{LB} has the form of the Yang-Baxter
relation, but in difference to  the $\mathcal R-$operator,
the operators ${\mathcal L}_{12}^\pm(\alpha)$ map the space
$\VV_{s_1}\otimes \VV_{s_2}\mapsto
\VV_{s_2\pm\alpha}\otimes \VV_{s_1\mp\alpha}$. However, for
the special values of the spectral parameter,
$\alpha_\pm=\pm(s_2-s_1)$, the operators ${\mathcal
L}_{12}^\pm(\alpha_\pm)$ coincide with the $\mathcal
R-$operator for the special values of the spectral
parameter,
\begin{equation}\label{LR}
{\mathcal L}_{12}^{\pm}(\pm(s_2-s_1))={\mathcal R}_{12}(\mp
i(s_2-s_1))\,,
\end{equation}
and play an important role in the subsequent construction.

%%%%%%%%%%%%%%%%%%%%%%%%%%%%%%%%%%%%%%%%%%%%%%%%%%%%%%%%%%%%%%%%%%%
%%%%%%%%%%%%%%%%%%%%%%%%%%%%%%%%%%%%%%%%%%%%%%%%%%%%%%%%%%%%%%%%%%%%
\begin{figure}[t]
\psfrag{w2}[cc][cc]{$\bar w_2$} \psfrag{w3}[cc][cc]{$\bar
w_3$} \psfrag{w5}[cc][cc]{$\bar w_1$}
\psfrag{w4}[cc][cc]{$\bar w_{N}$}
\psfrag{z1}[cc][cc]{$z_1$} \psfrag{z2}[cc][cc]{$z_2$}
\psfrag{z5}[cc][cc]{$z_3$} \psfrag{z4}[cc][cc]{$z_1$}
\psfrag{z3}[cc][cc]{$z_{N}$}
\psfrag{s1}[cr][cc]{$\balpha_u$}
\psfrag{s2}[cr][cc]{$\bbeta_u$}
\centerline{\includegraphics[scale=0.75]{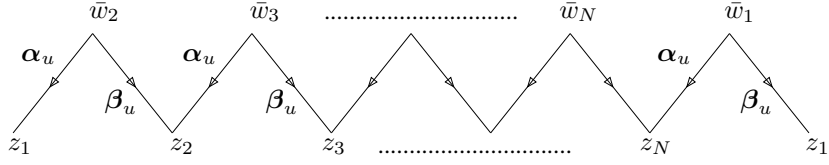}}
\caption[]{The graphical representation of  the kernel of
$Q(u)$ operator for the closed spin chain; $\alpha_u=s-iu$
and $\beta_u=s+iu$.} \label{Qcp}
\end{figure}
%%%%%%%%%%%%%%%%%%%%%%%%%%%%%%%%%%%%%%%%%%%%%%%%%%%%%%%%%%%%%%%%%%%%
%%%%%%%%%%%%%%%%%%%%%%%%%%%%%%%%%%%%%%%%%%%%%%%%%%%%%%%%%%%
\begin{figure}[t]
\psfrag{w1}[cc][cc]{$\bar w_1$}
\psfrag{w2}[cc][cc]{$\bar w_2$}
\psfrag{w3}[cc][cc]{$\bar w_3$}
\psfrag{w4}[cc][cc]{$\bar w_N$}

\psfrag{z1}[cc][cc]{$z_1$}
\psfrag{z2}[cc][cc]{$z_2$}
\psfrag{z3}[cc][cc]{$z_3$}

\psfrag{zn}[cc][cc]{$z_N $}

\psfrag{a}[cc][cc]{$\bgamma_u$}
\psfrag{s}[cc][rc]{$2s$}
\centerline{\includegraphics[scale=1.0]{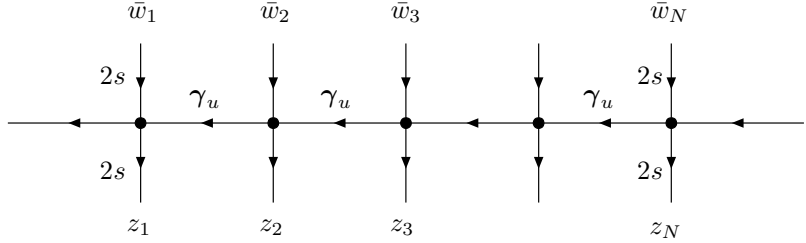}}
%\vspace*{0.5cm}
\caption[]{The kernel of the ${\widetilde Q}(u)$ operator
for the closed spin chain. All horizontal lines carry the
index
 $\gamma_u=iu-s$, while the vertical ones have the index $2s$.
 The black dots denote the integration
with the measure corresponding to the  spin $s'=(s+iu)/2$. }
\label{Qcm}
\end{figure}
%%%%%%%%%%%%%%%%%%%%%%%%%%%%%%%%%%%%%%%%%%%%%%%%%%%%%%%%%%%%%%%%%%%%

In what follows we show that the transfer matrix for the
closed homogeneous spin chain~\eref{To} can be represented
in the factorized form
\begin{equation}\label{Facc}\fl
{\mathbf T}_{s_0}^{\mathrm{cl}}(u)=Q(u+is_0)\,\widetilde Q(u-is_0)=
\widetilde Q(u-is_0)\,Q(u+is_0)\,,
\end{equation}
where $s\ell(2)$ invariant $Q$-operators are given by the
traces of ${\mathcal L}_{12}^\pm$ operators. Namely, we get
\begin{eqnarray}\label{QL}
Q(u)=\tr_{s_0} {\mathcal L}_{10}^-({s-s_0})\ldots
{\mathcal L}_{N0}^-(s-s_0)\bigl|_{s_0=(s-iu)/2}\,,\\[2mm]
\widetilde Q(u)={\mathcal P}\tr_{s_0} {\mathcal
L}_{10}^+(s_0-s)\ldots {\mathcal
L}_{N0}^+(s_0-s)\bigl|_{s_0=(s+iu)/2}\,,
\end{eqnarray}
where ${\mathcal P}$ is the cyclic permutation operator,
${\mathcal P}\psi(z_1,z_2,\ldots,z_N)=\psi(z_2,z_3,\ldots,z_1)$.
Taking into account \eref{LR} we conclude that the operators $Q(u)$ and $\widetilde
Q(u)$ coincide with the transfer matrices
\begin{eqnarray}
\label{QpT}
Q(u)={\mathbf T}_{(s-iu)/2}\left(\frac{u-is}{2}\right)\,,\ \ \ \ \\[2mm]
\label{QmT}
\widetilde Q(u)={\mathcal P}{\mathbf T}_{(s+iu)/2}\left(\frac{u+is}{2}\right)\,.
\end{eqnarray}
The commutativity of the $Q$, $\widetilde Q$ operators,
 $[Q(u) , Q(v)]=[\widetilde Q(u) , \widetilde Q(v)]=
 [Q(u) , \widetilde Q(v)] = 0$,
follows immediately from the commutativity of the transfer
matrices, (we remind that ${\mathcal P}^{-1}={\mathbf
T}^{s}_{s}(0)$.)

Making use of Eqs.~\eref{QpT},~\eref{QmT} one can represent
Eq.~\eref{Facc} in the following form
\begin{equation}\fl
{\mathbf T}_{s_0}(u)={\mathbf T}_{(s+s_0-iu)/2}\left(\frac{u-i(s-s_0)}{2}\right)\,
{\mathcal P}\,{\mathbf T}_{(s+s_0+iu)/2}\left(\frac{u+i(s-s_0)}{2}\right)\,.
\end{equation}

\begin{figure}[t]
\psfrag{w1}[cc][cc]{$\bar w_1$}
\psfrag{w2}[cc][cc]{$\bar w_2$}
\psfrag{w3}[cc][cc]{$\bar w_3$}
\psfrag{wn}[cc][cc]{$\bar w_N$}

\psfrag{z1}[cc][cc]{$z_1$}
\psfrag{z2}[cc][cc]{$z_2$}
\psfrag{z3}[cc][cc]{$z_3$}

\psfrag{zn}[cc][cc]{$z_N $}
\psfrag{eq}[cc][bc]{$=$}
\psfrag{a}[cc][cc]{$\alpha_u$}
\psfrag{b}[cc][cc]{$\beta_u$}
\psfrag{c}[cc][cc]{$\gamma_u$}
\psfrag{s}[cc][cc][0.8]{$2s$}
\centerline{\includegraphics[scale=0.62]{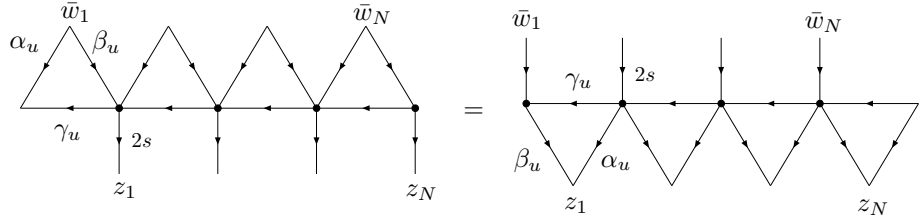}}
%\vspace*{0.5cm}
\caption[]{The graphical representation of the transfer matrix for the closed spin chain.
The indices $\alpha_u=s-i(u+is_0)$, $\beta_u=s+i(u+is_0)$. The indices
of the vertical lines are equal to $2s$, and those of the horizontal are equal to
$\gamma_u=i(u-is_0)-s$. The black dots denotes the integration vertices
corresponding to the  spin $s'=(s+i(u-is_0))/2$.}
\label{trcl}
\end{figure}

To prove the factorization property  we shall show that the
integral kernels of the operators on the l.h.s and r.h.s of
Eq.~\eref{Facc} coincide. To this end let us   represent
the kernel of the  operators under consideration in the
graphical form. The diagrammatical representation of the
kernels for the operators $Q(u)$ and $\widetilde Q(u)$ are
shown in Figures~\ref{Qcp} and \ref{Qcm}, respectively. In
its turn, the integral kernel for the transfer matrix  is
shown, in two equivalent forms, in Figure~\ref{trcl}.
%%%%%%%%%%%%%%%%%%%%%%%%%%%%%%%%%%%%%%%%%%%%%%%%%%%%%%%%%%%%%%%%%%%%
%%%%%%%%%%%%%%%%%%%%%%%%%%%%%%%%%%%%%%%%%%%%%%%%%%%%%%%%%%%%%%%%%%%%
%%%%%%%%%%%%%%%%%%%%%%%%%%%%%%%%%%%%%%%%%%%%%%%%%%%%%%%%%%%%%%%%%%%%
%%%%%%%%%%%%%%%%%%%%%%%%%%%%%%%%%%%%%%%%%%%%%%%%%%%%%%%%%%%%%%%%%%%%
Drawing the diagram for the product $Q(u+is_0) \widetilde Q(u-is_0)$
(or $\widetilde Q(u-is_0) Q(u+is_0)$) one notices that the measure of  integration
in the intermediate triple vertices corresponds to the spin $s$. Since the
vertical lines attached to this vertex correspond to the reproducing kernel with the spin $s$,
one can carry out the integration using
the property~\eref{KK} and  find that the resulting  diagram coincides with
the diagram for the kernel of the transfer matrix.
Thus the property of the factorization for  the
homogenous spin chain is established.

For completeness, we write down the analytic expressions for the kernels of the
$Q-$operators,
\begin{eqnarray}\fl\label{Qucp}
Q(u)(\Mybf{z}|\Mybf{w})=\prod_{k=1}^N (1-z_k\bar w_k)^{-s-iu}(1-z_k\bar w_{k+1})^{-s+iu}\,,
\\[2mm]
\fl\label{Qucm} \widetilde Q(u)(\Mybf{z}|\Mybf{w})=
\prod_{k=1}^{N}
\int {\mathcal D}_{s'}\zeta_k
\,(1-\zeta_k\bar\zeta_{k+1})^{s-iu}(1-z_k\bar\zeta_k)^{-2s}\,
(1-\zeta_k\bar w_k)^{-2s}\,,
\end{eqnarray}
where $s'=s+iu/2$, and $w_{N+1}\equiv w_1$ and so on. The expression~\eref{Qucp} coincides
with the expression for the Baxter operator obtained in~Ref.~\cite{SD}.

\vskip 0.5cm

Let us consider now the  homogeneous $s\ell(2)$ invariant
open spin chain. The transfer matrix for the open spin
chain,~\eref{To}, can also  be represented in the factorized
form, namely
\begin{equation}\fl\label{TQQ}
{\mathbf T}_{s_0}^{\mathrm{op}}(u)=g(u)
{\mathcal Q}(u+is_0)\,\widetilde {\mathcal Q}(u-is_0)=g(u)
\widetilde{\mathcal  Q}(u-is_0)\,{\mathcal Q}(u+is_0)\,,
\end{equation}
where
\begin{equation}\label{gu}
g(u)=\frac{s+s_0+iu-1}{2iu-1}\,.
\end{equation}
The operators ${\mathcal Q}(u)$ and $\widetilde{\mathcal
Q}(u)$ have the following form
\begin{eqnarray}\label{QQL} \fl
{\mathcal Q}(u)=\tr_{s_0} {\mathcal
L}_{10}^-({s-s_0})\ldots {\mathcal
L}_{N0}^-(s-s_0){\mathcal L}_{N0}^-(s-s_0)\ldots {\mathcal
L}_{10}^-({s-s_0})
\bigl|_{s_0=\frac{s-iu}{2}}\,,\\[2mm]
\fl \widetilde {\mathcal Q}(u)=\tr_{s_0} {\mathcal
L}_{10}^+(s_0-s)\ldots {\mathcal L}_{N0}^+(s_0-s){\mathcal
L}_{N0}^+(s_0-s)\ldots{\mathcal L}_{10}^+(s_0-s)
\bigl|_{s_0=\frac{s+iu}{2}}\,.
\end{eqnarray}
Again, taking into account  Eq.~\eref{LR} one relates $\mathcal Q-$operators to the transfer
matrices for the open spin chain
\begin{eqnarray}
\label{TpQ}
{\mathcal Q}(u)={\mathbf T}_{(s-iu)/2}\left(\frac{u-is}{2}\right)\,,\ \ \ \ \\[2mm]
\label{TmQ}
\widetilde {\mathcal Q}(u)={\mathbf T}_{(s+iu)/2}\left(\frac{u+is}{2}\right)\,.
\end{eqnarray}
Thus, similarly to the closed spin chain, one concludes that $\mathcal Q-$operators
commute with each other for arbitrary values of the spectral parameters.
 %%%%%%%%%%%%%%%%%%%%%%%%%%%%%%%%%%%%%%%%%%%%%%%%%%%%%%%%%%%
\begin{figure}[t]
\psfrag{z1}[cc][cc]{$z_1$}\psfrag{z2}[cc][cc]{$z_2$}\psfrag{zn}[cc][cc]{$z_N$}
\psfrag{w1}[cc][cc]{$\bar w_1$}\psfrag{w2}[cc][cc]{$\bar w_2$}
\psfrag{wn}[cc][cc]{$\bar w_N$} \psfrag{dots}[cc][cc]{$\Mybf{\cdots}$}
\psfrag{a}[cc][cc]{$\beta_u$} \psfrag{b}[cc][cc]{$\alpha_u$}
\psfrag{c}[cc][rc]{$\alpha_u$} \psfrag{d}[cc][rc]{$\beta_u$}
\centerline{\includegraphics[scale=0.8]{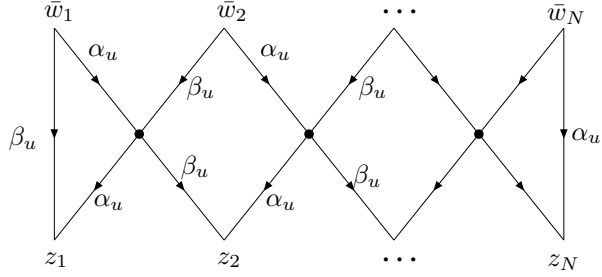}} %\vspace*{0.5cm}
\caption[]{Diagrammatical representation of the ${\mathcal
Q}(u)-$operator for the open spin chain;
$\alpha_u=s-iu$ and $\beta_u=s+iu$.}%
\label{QO+}%
\end{figure}%
%%%%%%%%%%%%%%%%%%%%%%%%%%%%%%%%%%%%%%%%%%%%%%%%%%%%%%%%%%%%%%%%%%%%
%%%%%%%%%%%%%%%%%%%%%%%%%%%%%%%%%%%%%%%%%%%%%%%%%%%%%%%%%%%
\begin{figure}[t]
\psfrag{z1}[cc][cc]{$z_1$}
\psfrag{z2}[cc][cc]{$z_2$}
\psfrag{zn}[cc][cc]{$z_N$}
\psfrag{w1}[cc][cc]{$\bar w_1$}
\psfrag{w2}[cc][cc]{$\bar w_2$}
\psfrag{wn}[cc][cc]{$\bar w_N$}
%\psfrag{dots}[cc][cc]{$\Mybf{\cdots}$}
%\psfrag{a}[cc][cc]{$\beta_u$} \psfrag{b}[cc][cc]{$\alpha_u$}
\psfrag{c}[cc][cc]{$\gamma_u$} \psfrag{d}[cc][rc]{$\beta_u$}
\centerline{\includegraphics[scale=0.8]{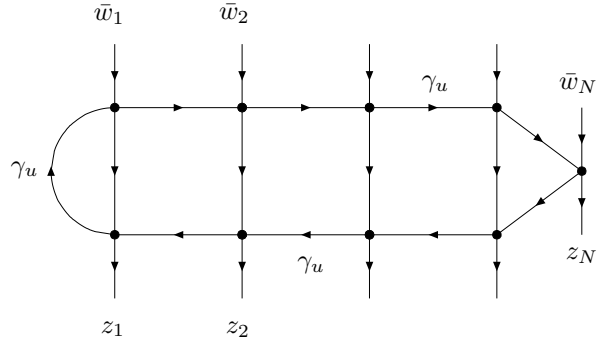}} %\vspace*{0.5cm}
\caption[]{Diagrammatical representation of the
${\widetilde{\mathcal Q}(u)}-$operator for the open spin
chain. Here the all horizontal lines carry the index
 $\gamma_u=iu-s$, all vertical lines have index $2s$.
 The black dots denote the integration
with the measure corresponding to the conformal spin $s'=(s+iu)/2$.
}%
\label{QO-}%
\end{figure}%
%%%%%%%%%%%%%%%%%%%%%%%%%%%%%%%%%%%%%%%%%%%%%%%%%%%%%%%%%%%%%%%%%%%%

\begin{figure}[t]
\psfrag{w1}[cc][cc]{$\bar w_1$}
\psfrag{w2}[cc][cc]{$\bar w_2$}
\psfrag{w3}[cc][cc]{$\bar w_3$}
\psfrag{wn}[cc][cc]{$\bar w_N$}

\psfrag{z1}[cc][cc]{$z_1$}
\psfrag{z2}[cc][cc]{$z_2$}
\psfrag{z3}[cc][cc]{$z_3$}
\psfrag{gu}[cc][cc]{$g(u)\times$}

\psfrag{zn}[cc][cc]{$z_N $}
\psfrag{2u}[cc][bc][0.9]{$2iu$}
\psfrag{a}[cc][cc]{$\alpha_u$}
\psfrag{b}[cc][cc]{$\beta_u$}
\psfrag{c}[cc][cc]{$\gamma_u$}
\psfrag{2s}[cc][cc][0.8]{$2s$}
\centerline{\includegraphics[scale=0.68]{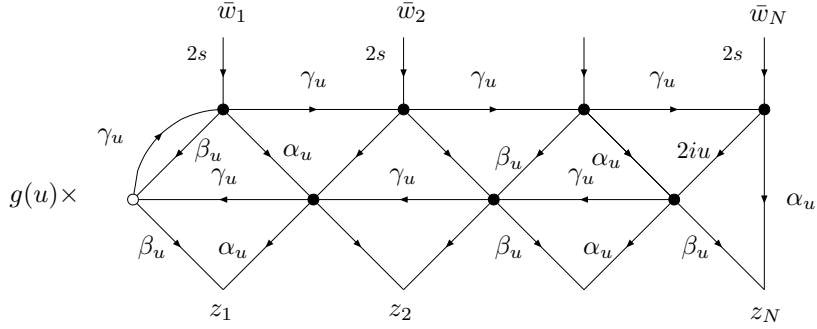}}
%\vspace*{0.5cm}
\caption[]{The graphical representation of the transfer matrix for the open spin chain.
Here $\alpha_u=s-i(u+is_0)$, $\beta_u=s+i(u+is_0)$, $\gamma_u=i(u-is_0)-s$,
The black dots denotes the integration vertices
corresponding to the  spin $s'=(s+i(u-is_0))/2$ and white dot denotes the integration
vertex corresponding to the spin $s''=iu$. The prefactor $g(u)$ is
given by  Eq.~\eref{gu}.
\label{trop}}
\end{figure}

To prove the factorization~\eref{TQQ} we again use the
graphical representation for kernels.  For the ${\mathcal
Q(u)}$ and $\widetilde {\mathcal Q}(u)$ operators they are
shown in Figure~\ref{QO+} and~\ref{QO-}, respectively.  The
diagrammatical representation for the kernel of the
transfer matrix for the open spin chain is shown in
Figure~\ref{trop}.  In order to derive this representation one starts with the
definition~\eref{To} and uses the graphical representation
for the kernel of $\mathcal R-$operator shown on the lhs of
Figure~\ref{RR}. Next one should carry out the integration
over all intermediate ``quantum'' vertices
$\zeta_1,\ldots,\zeta_N$. The integration measure in each
vertex is given by the expression~\eref{DS} where $s$ is
the  ``spin'' of  the quantum space. The result of the
integration is the disappearance of the lines with index
$2s$
\begin{equation}\fl\label{example}
\int D_s'\xi \int{\mathcal  D}\zeta_s \psi(\xi,\bar \xi){\mathbb K}_s(\xi,\zeta)\phi(\zeta)=
\int D_s'\xi \psi(\xi,\bar \xi)\phi(\xi)\,
\end{equation}
attached to these vertices.
Finally one can carry out the integration over
``auxiliary space'' vertices. Again, the lines attached to these vertices disappear.
The line with the index $2iu$ (on the right part of the diagram)
arises due to merging of  two lines with indices $\gamma_u$ and
$\beta_u$, $(1-z\bar w)^{-\gamma_u}(1-z\bar w)^{-\beta_u}=(1-z\bar w)^{-2iu}$.

To explain the appearance of the factor $g(u)$ and the integration vertex with spin
$s''=iu$, we notice that after the integration one line with index $\alpha_u$
becomes attached to the vertex with the
spin $s'=s+i(u-is_0))$ by both ends. Noticing that
$$
\int D_{s'}\xi (1-\xi\bar \xi)^{-\alpha_u}\ldots=g(u)\int D_{iu}\xi\ldots
$$
one obtains finally  the diagrammatic representation  for
the kernel of the transfer matrix shown in
Figure~\ref{trop}.
%%%%%%%%%%%%%%%%%%%%%%%%%%%%%%%%%%%%%%%%%%%%%%%%%%%%%%%%%%%%%%%%%%%%
%%%%%%%%%%%%%%%%%%%%%%%%%%%%%%%%%%%%%%%%%%%%%%%%%%%%%%%%%%%%%%%%%%%%
%%%%%%%%%%%%%%%%%%%%%%%%%%%%%%%%%%%%%%%%%%%%%%%%%%%%%%%%%%%%%%%%%%%%
%%%%%%%%%%%%%%%%%%%%%%%%%%%%%%%%%%%%%%%%%%%%%%%%%%%%%%%%%%%%%%%%%%%%

\begin{figure}[t]
\psfrag{w1}[cc][cc]{$\bar w_1$}
\psfrag{w2}[cc][cc]{$\bar w_2$}
\psfrag{w3}[cc][cc]{$\bar w_3$}
\psfrag{wn}[cc][cc]{$\bar w_N$}

\psfrag{z1}[cc][cc]{$z_1$}
\psfrag{z2}[cc][cc]{$z_2$}
\psfrag{z3}[cc][cc]{$z_3$}
\psfrag{gu}[cc][cc]{}

\psfrag{zn}[cc][cc]{$z_N $}
\psfrag{2u}[cc][bc][0.8]{$2iu$}
\psfrag{a}[cc][cc][0.9]{$\alpha_u$}
\psfrag{b}[cc][cc][0.9]{$\beta_u$}
\psfrag{c}[cc][cc][0.9]{$\gamma_u$}
\psfrag{2s}[cc][cc][0.8]{$2s$}
\centerline{\includegraphics[scale=0.63]{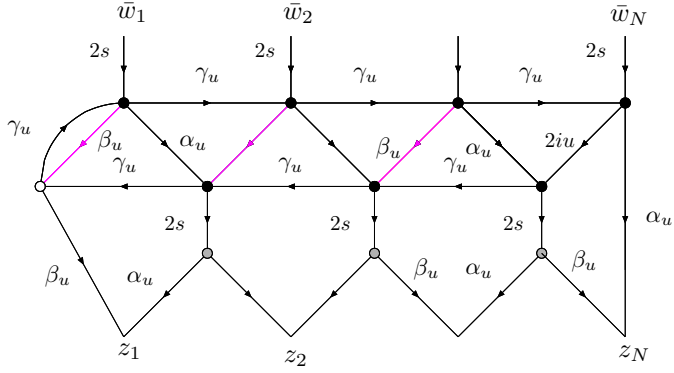}}
%\vspace*{0.5cm}
\caption[]{The diagram for the kernel of the transfer matrix for the open spin chain.
All notations the same as in Figure \ref{trop}. The gray dots denote the integration
vertices corresponding to the  spin $s$.
\label{transf-1}}
\end{figure}

Now we have to show that the diagram for the transfer
matrix can be transformed to the diagram for the product of
the operators ${\mathcal Q}(u+is_0)\widetilde {\mathcal
Q}(u-is_0)$. The first transformation is the insertion of
the reproducing kernels into the diagram in Fig.~\ref{trop}
as shown in Figure~\ref{transf-1}. This operation does not
change the kernel, since after the integration
over the new vertices one reproduces the initial
expression. The next transformation is the following. Let
us consider the subdiagram formed by the four lines (which have
indices $2iu, \alpha_u,\gamma_u, 2s$) attached to the right
(black) vertex in the middle line of the transformed
diagram, and the line with index $\beta_u$ which connects
the lines with indices $\alpha_u$ and $\gamma_u$. It can be
checked that the indices satisfy the conditions
$\alpha_u+2iu=2s'=\gamma_u+2s$ and
$\beta_u=2s'-\alpha_u-\gamma_u$. It allows one to use the
permutation relation shown in Figure~\ref{RR}. After the 
transformation, the line with the index $\beta_u$ changes its
position and will connect the endpoints of the other pair
of lines. In addition, the indices of the lines in the new
diagrams have to be changed, namely one should interchange
$\alpha_u$ and $2s$ (~$\alpha_u\leftrightarrow 2s$~) and
$\gamma_u$ and $2iu$ (~$\gamma_u\leftrightarrow 2iu$~).

Next one notices that the subdiagram formed by the lines
attached to the next vertex has exactly the same form as
the one considered just now. Therefore one can repeat
this  transformation successively. As the result all lines with index
$\beta_u$ in the upper part of the diagram change their
positions, and one has also to interchange the indices in
the way described above, namely, 
$\alpha_u\leftrightarrow 2s$. Further, since  the
interchange $2iu\leftrightarrow\gamma_u$ occurs twice for
the all lines except for the first and the last one in this
chain,  the whole effect will be that the line attached to
the leftmost vertex (white blob) in the figure will get the
index $2iu$, while all other ``horizontal'' lines will have
the same index, $\gamma_u$.

It is important  that after this series of 
transformations only three lines will be attached to the
leftmost vertex (white blob in the Figure~\ref{transf-1}).
The integration measure in this vertex corresponds to the
conformal spin $s''=iu$. Since the incoming arrow has 
index $2iu$,
 and two other arrows come out of the vertex, one can integrate
over this vertex. After the integration 
the line with index $2iu$ disappears (see
Eq.~\eref{example}) and one can easily check that the
resulting diagram has the form of the integral kernel for
the operator
 ${\mathcal Q}(u+is_0)\widetilde {\mathcal Q}(u-is_0)$.
%%%%%%%%%%%%%%%%%%%%%%%%%%%%%%%%%%%%%%%%%%%%%%%%%%%%%%%%%%%%%%%%%%%%
%%%%%%%%%%%%%%%%%%%%%%%%%%%%%%%%%%%%%%%%%%%%%%%%%%%%%%%%%%%%%%%%%%%%
%%%%%%%%%%%%%%%%%%%%%%%%%%%%%%%%%%%%%%%%%%%%%%%%%%%%%%%%%%%%%%%%%%%%
%%%%%%%%%%%%%%%%%%%%%%%%%%%%%%%%%%%%%%%%%%%%%%%%%%%%%%%%%%%%%%%%%%%%

\setcounter{equation}{0}
\section{Baxter equation}\label{BAX}
In this section we study the relation between $Q-$operators and the transfer matrices
over finite dimensional auxiliary spaces.
For a negative (half)integer values of the spin of auxiliary space, $s_0=-n$,
the representation space $\VV_{s_0=-n}$ contains an invariant subspace $V_{n}$.
Thus, the subspace $\VV_s\otimes V_{n}$ is an invariant subspace of the
${\mathcal R}_{ss_0}-$operator. It has, therefore, the triangular form
\begin{equation}\label{triang}
{\mathcal R}_{ss_0}(u)=\left(\begin{array}{cc}  r_{ss_0}(u)& *\\
0&\widetilde{\mathcal R}_{ss_0}(u)
\end{array}
\right)\,,
\end{equation}
where $r_{ss_0}(u)$ is the restriction of the operator
${\mathcal R}_{ss_0}$ to the subspace $\VV_s\otimes V_{n}$.
The operator $\widetilde{\mathcal R}_{ss_0}(u)$ acts on the
space $\VV_s\otimes \VV_{s_0}/V_{n}\sim\VV_s\otimes
\VV_{1+n/2}$ and satisfies the YB relation~\eref{YB}.
Therefore it has to be proportional to ${\mathcal
R}_{s,s'_0}(u)$ with the spin $s'_0=1-s_0=1+n$
\begin{equation}\label{RtoR}
\widetilde {\mathcal R}_{s,s_0=-n}(u)=f_n(u){\mathcal
R}_{s,s_0=1+n}(u)\,.
\end{equation}
The normalization coefficient
\begin{equation}
f_n(u)=(-1)^{2n+1}\frac{\Gamma(s+1+n-iu)}{\Gamma(s-n-iu)}
\frac{\Gamma(s-n+iu)}{\Gamma(s+1+n+iu)}\,
\end{equation}
can be found by comparing  the eigenvalues of
$\widetilde{\mathcal R}_{ss_0}(u)$ and ${\mathcal
R}_{s,s'_0}(u)$ on the eigenstates,
$\psi_k(z_1,z_2)=(z_1-z_2)^k$, or by using  the
intertwining relation for the $\mathcal{R}^\pm$-operators. The latter takes the form
\begin{eqnarray}
\frac{1}{z_{01}^{2n+1}}\cdot
\frac{\Gamma(z_{01}\dd_0+2(s_0-\alpha))}
{\Gamma(z_{01}\dd_0+2s_0)} &=&
\frac{\Gamma(z_{01}\dd_0+2(s'_0-\alpha'))}
{\Gamma(z_{01}\dd_0+2s'_0)}\dd_0^{2n+1} \ ,\
\\[2mm]
\dd_1^{2n+1}\frac{\Gamma(z_{10}\dd_1+2s_0)}
{\Gamma(z_{10}\dd_1+2(s_0-\alpha))} &=&
\frac{\Gamma(z_{10}\dd_1+2s'_0)}
{\Gamma(z_{10}\dd_1+2(s'_0-\alpha'))}\frac{1}{z_{10}^{2n+1}}
\ ,
\end{eqnarray}
where $ s_0 =-n$, $s'_0=1+n$, $\alpha'=\alpha+n+1/2$, $n$ being half-integer.

Equations~\eref{triang},~\eref{RtoR} imply the following
relation for the transfer matrices of the closed spin
chain
\begin{equation}\label{TTc}
{\mathbf T}_{s_0=-n}(u)=t_{n}(u)+(f_n(u))^N {\mathbf T}_{s_0=1+n}(u)\,,
\end{equation}
where $t_{n}(u)$ is the transfer matrix with finite
dimensional auxiliary space,  \mbox{$t_{n}(u)=\tr_{V_{n}} r_{10}(u)\ldots r_{N0}(u)$}.
Using the factorization property~\eref{Facc} and introducing the notation
\begin{equation}\label{QQt}
\Delta(u)\,{\widetilde Q}(u)=
\widebar Q(u+i)\,,
\end{equation}
where
\begin{equation}
\Delta(u)=\left(\frac{\Gamma(1-s-iu)}{\Gamma(1+s-iu)}\right)^N
\end{equation}
we rewrite Eq.~\eref{TTc} in the form
\begin{equation}\fl\label{tQQ}
\Delta(u+in)\,t_{n}(u)=Q(u-in)\, \widebar
Q(u+i(1+n))-Q(u+i(1+n))\, \widebar Q(u-in)\,.
\end{equation}
For $n=0$ the transfer matrix $t_{n=0}(u)=1$ and
Eq.~\eref{tQQ} reads
\begin{equation}%\fl
\left(\frac{\Gamma(1-s-iu)}{\Gamma(1+s-iu)}\right)^N=
Q(u)\,\widebar Q(u+i)-Q(u+i)\,\widebar Q(u)\,,
\end{equation}
which is the Wronskian relation between the operators $Q$ and $ \widebar Q$.
Further, one can exclude the operator $\widebar Q$ (or $Q$) from eq.~\eref{tQQ}.
Indeed, multiplying both sides by $Q(u-im)$, where $m$ is (half)integer such that
$m+n$ is integer and $-n\leq m\leq n-1$, after some algebra one finds
\begin{eqnarray}\fl\label{3tQ}
\tilde t_n(u)\,Q(u-im)&=&\tilde t_{(n+m)/2}\left(u+\frac{i(n-m)}{2}\right)\,Q(u-in)+\nonumber\\[2mm]
\fl&&\tilde t_{(n-m-1)/2}\left(u-\frac{i(n+m+1)}{2}\right)\,Q(u+i(n+1))\,,
\end{eqnarray}
where $\tilde t_n(u)=\Delta(u+in)t_n(u)$. The same equation
holds for the $\widebar Q$ operator as well. The first relation
($n=-m=1/2$) among the ones in~\eref{3tQ} is nothing else but
the Baxter equation
\begin{equation}\label{BE}
\tau_N(u)\,Q(u)=(u+is)^N\,Q(u+i)+(u-is)^N\,Q(u-i)\,,
\end{equation}
where
$$
\tau_N(u)=\tr L_1(u)\ldots L_N(u)
$$
and $L(u)$ is the Lax operator
$$
L(u)=u+i\left(\begin{array}{cc}S_0&S_-\\S_+&-S_0\end{array}\right).
$$
To derive Baxter equation~\eref{BE} from \eref{3tQ} one put
$n=-m=1/2$ and takes into account that $t_0(u)=1$ and
$t_{1/2}(u-i/2)=(u-is)^{-N}\tau_N(u)$. The latter relation follows
from the relation $L(u)=(u-is)\, r_{s,-1/2}(u-i/2)$ which
can easily be  verified by comparison of the
eigenvalues~\cite{KRS}. It is evident that the operator
$\widebar Q(u)$ satisfies the same equation~\eref{BE}. Thus
the operators $Q(u)$ and $\widebar Q(u)$ (and, as
consequence, their eigenvalues) represent two independent
solutions of the Baxter equation~\eref{BE}. It can be shown
\cite{SD} that the eigenvalues of the operator $Q(u)$ are
polynomials in $u$. The eigenvalues of the second operator
$\widebar Q(u)$ are meromorphic functions of $u$ with poles
of  order $N$ ($N$ is the length of the chain) at the
points $u_k=-i(1-s+k)$, $k=0,1,2\ldots,\infty$.

Next, using the triple relation~\eref{3tQ} one can derive
the "fusion" relations for the transfer matrices. Indeed,
multiplying the both sides of  Eq.~\eref{3tQ} by the
operator $\widetilde Q(v)$ and using the factorized
expression for the transfer matrix~\eref{Facc} one obtaines
the relation which involves three $t$ and $T$ transfer matrices.
After the substitution $i(v-u)/2\to s_0$ and $(u+v)/2\to u$
it takes  form
\begin{eqnarray}\fl
\label{fusion-1}
{ t}_n(u+is_0)\,T_{s_0-\frac{m}2}\left(u-\frac{im}2\right)&=&
{ t}_{\frac{n+m}2}\left(u+is_0+\frac{i(n-m)}2\right)\,
T_{s_0-\frac{n}2}\left(u-\frac{in}2\right)+\nonumber\\[2mm]
&&\hskip -2.0cm f_{nm}(u+is_0)\,
{ t}_{\frac{n-m-1}2}\left(u+is_0-\frac{i(n+m+1)}2\right)\,
T_{s_0+\frac{n+1}2}\left(u+\frac{i(n+1)}2\right)\,,
\end{eqnarray}
with
\begin{equation}
 f_{nm}(u)=\frac{\Delta(u-i(m+1))}{\Delta(u+in)}\,.
\end{equation}
Similarly, starting from Eq.~\eref{3tQ} involving $\widebar
Q$ operator and multiplying by $Q(v)$ one gets another
identity
\begin{eqnarray}\fl\label{fusion-2}
{ t}_n(u-is_0)\, T_{s_0+\frac{m+1}2}\left(u-\frac{i(m+1)}2\right)&=&
{ t}_{\frac{n-m-1}2}\left(u-is_0-\frac{i(n+m+1)}2\right)\,
T_{s_0-\frac{n}2}\left(u+\frac{in}2\right)+\nonumber\\[2mm]
&&\hskip -2.7cm f_{-m-1,n}(u-is_0)\,{t}_{\frac{n+m}2}\left(u-is_0+\frac{i(n-m)}2\right)\,
T_{s_0+\frac{n+1}2}\left(u-\frac{i(n+1)}2\right)\,.
\end{eqnarray}
For $n=-m=1/2$ the relations~\eref{fusion-1},~\eref{fusion-2} take the standard
form~\cite{KRS}
and relate the transfer matrices with adjacment spins of auxiliary space, $T_{s_0}$ and
$T_{s_0\pm 1/2}$.

Next, starting from the Eqs.~\eref{tQQ} and \eref{3tQ} one can derive two quadratic
relations for the finite-dimensional transfer matrices $t_n(u)$. The first one is 
\begin{eqnarray}\fl\label{tt2}
\tilde t_{\frac{m-n-1}2}\left(u+\frac{ik}{2}\right)\tilde
t_{\frac{m+n-1}2}\left(u-\frac{ik}{2}\right)&=&
\tilde t_{\frac{m-k-1}2}\left(u+\frac{in}{2}\right)\tilde
t_{\frac{m+k-1}2}\left(u-\frac{in}{2}\right)\nonumber\\
&&+
\tilde t_{\frac{k-n-1}2}\left(u+\frac{im}{2}\right)\tilde
t_{\frac{k+n-1}2}\left(u-\frac{im}{2}\right)\,.	
\end{eqnarray}	
Here the numbers $m,k,n$ are all integer or half-integer  and $m>k>n$. The second
relation is obtained from the first one by  changing $n\to -n$.

\vskip 0.5cm

The treatment of the open spin chain goes  along the same lines.
The analog of Eq.~\eref{tQQ} reads
\begin{equation}\fl\label{tQQo}
(2iu-1)\,\Delta_n^2(u)\,t_{n}^{\mathrm{op}}(u)={\mathcal Q}(u-in)\, \widebar{\mathcal  Q}(u+i(1+n))-
{\mathcal Q}(u+i(1+n))\,
\widebar{\mathcal Q}(u-in)\,,
\end{equation}
where
\begin{equation}
\widetilde {\mathcal Q}(u)=\frac{1}{s-i+iu}\left(\frac{\Gamma(1+s-iu)}{\Gamma(1-s-iu)}\right)^{2N}
\widebar {\mathcal Q}(u+i)
\end{equation}
and
$$
t_{n}^{\mathrm{op}}(u)=\tr_{V_{n}} r_{10}(u)\ldots r_{N0}(u)\,r_{N0}^{-1}(-u) \ldots r_{10}^{-1}(-u).
$$
Obviously,  the operator ${\mathcal Q}(u)$ ($\widebar
{\mathcal Q}(u)$) for the open chain satisfies the same
Eq.~\eref{3tQ} with $\tilde t_n(u)=(2iu-1)\,\Delta^2_n(u)\,
t_{n}^{\mathrm{op}}(u)$. The Wronskian relation and Baxter
equation take the well known form~ \cite{ES88,open}
\begin{eqnarray}\fl
&&(2iu-1)\left(\frac{\Gamma(1-s-iu)}{\Gamma(1+s-iu)}\right)^{2N}=
{\mathcal Q}(u)\,\widebar {\mathcal Q}(u+i)-{\mathcal Q}(u+i)\,\widebar{\mathcal Q}(u)\,,\\[2mm]
\fl
&&\tau_N^{\mathrm{op}}(u){\mathcal Q}(u)=\frac{2iu+1}{2iu}(u+is)^{2N}{\mathcal Q}(u+i)+
\frac{2iu-1}{2iu}(u-is)^{2N}{\mathcal Q}(u-i)\,.
\end{eqnarray}
\vskip  0.2mm \noindent Here $\tau_N^{\mathrm{op}}(u)=\tr
L_1(u)\ldots L_N(u)\,L_N(u)\ldots L_1(u)$. Again, the
eigenvalues of the operator $Q(u)$ are polynomials in
$u$~\cite{open}, while the eigenvalues of $\widebar Q$ are
meromorphic functions.

In full analogy with the closed spin chain one can derive
two sets of the "fusion" relations and check that the finite-dimensional transfer
matrices, $\tilde t_n(u)=(2iu-1)\,\Delta^2_n(u)\,
t_{n}^{\mathrm{op}}(u)$, satisfy the quadratic relation~\eref{tt2}.

\vskip 0.3cm

\section{Conclusions}\label{Concl}
In this paper we considered the quantum spin chains with
$s\ell(2)$ symmetry.  The Hilbert space of the model is given
by the tensor product of the $s\ell(2)$ modules. For a generic
spin $s$, the latter are infinite dimensional and
equivalent to the space of the polynomials of an arbitrary
degree, $\mathbf{V}_s=\mathbb{C}[z]$.
Using the factorization
of the $\mathcal R$ operator 
$$
{\mathcal R}_{12}(u) = P_{12}\, {\mathcal R}^{+}_{12}\left(\frac{s_2-s_1+iu}{2}\right)\, 
{\mathcal R}^{-}_{12}\left(\frac{s_1-s_2+iu}{2}\right)
$$
we have shown that the
transfer matrices both for the closed and open homogeneous
spin chains (for generic spin of the auxiliary space $s_0$)
factorize into a product of two commuting operators $Q$ and
$\widetilde Q$. 
The latter are given by the trace of the product of operators 
${\mathcal L}^\pm_{k0}={\mathcal R}_{k0}(\mp i(s_0-s))$ over the auxiliary space
(see Eqs.~\eref{QL},~\eref{QQL}).

For  negative half-integer spins,
$s_0=-n$, $n=0,1/2,1,\ldots$, the module $\mathbf{V}_s$ has
a finite dimensional invariant subspace, ${V}_{n}$.
The representation induced on the factor space
$\VV_{-n}/V_{n}$ is equivalent to the $s\ell(2)$ module with
spin $s'_0=1+n$. We have shown, that the operators $Q$ and
$\widetilde Q$ satisfy the finite-difference (Baxter)
equation, which follows unambiguously from the structure of
the reducible $s\ell(2)$ modules and factorization property of
the transfer matrices. The treatment of the closed and open
spin chains goes along the same lines with minor
differences. We hope that similar analysis will be
applicable to the spin chains with higher rank symmetry
groups.

The factorization property  of the transfer matrices breaks
down for the finite dimensional spin chain, therefore they
require a special consideration. The problem of
construction of the Baxter $Q$ operator for the finite
dimensional spin chains (in particular $XXX_{1/2}$ spin
magnet) was considered in the Refs.~\cite{Pronko,Korff}.

\ack
The authors are grateful to G.~P.~Korchemsky for helpful discussions.
This work was supported by the RFFI grants 03-01-00837 (S.~D. and A.~M.) 
and 05-01-0092 (S.~D.) 
and by the Helmholtz Association, contract number VH-NG-004 (A.~M.)

%%%%%%%%%%%%%%%%%%%%%%%%%%%%%%%%%%%%%%%%%%%%%%%%%%%%%%%%%%%%%%%%%%%%%%%%%%%%%%%%%%%%%%%%%%%%%

\end{document}